\documentclass[12pt]{article}
\usepackage{graphicx}
\usepackage{latexsym}

\begin{document}

\title{Quantum Bouncer: Theory and Experiment}
\author{Anatoli Andrei Vankov\\         
{\small \it IPPE, Obninsk, Russia; Bethany College, KS, USA,  anatolivankov@hotmail.com}}

\date{}

\maketitle

\begin{abstract}

The quantum bouncer (QB) concept is a known QM textbook example of confined particle, namely, a solution to the 1D Schroedinger equation for a linear potential (the so-called Airy equation). It would be a great methodological challenge to create such a QM object in laboratory. An attempt of observation of the QB ``running'' in the horizontal direction was recently made  by the international team at the Laue-Langevin Institute, Grenoble. The experiment was performed with ultra-cold neutrons. In this paper, the experiment is analyzed in view of the authors' claim that ``neutron quantum states in Earth gravitational field'' are observed. The experimental apparatus is designed for measurements of horizontal flux of neutrons passing through an absorbing wave guide with a variable height of absorber. From our analysis, it follows, however, that in such a layout measured data are not sensitive to quantum probability density in the vertical direction. The overall conclusion is made that the experimental data do not contain sufficient information to justify the claim. 
 
\medskip

Key words: quantum mechanics; ultracold neutrons; neutron quantum states; Earth gravitational field; experiment; Laue-Langevin Institute; Grenoble.

\
{\small\it PACS 04.80 Cc}
 
\end{abstract}

\section{Introduction}

There are textbook examples of particular solutions to the 1D  Schroedinger equation in terms of variable wavenumber $k(z)$) for a particle confined in a potential well $W(z)$ (the latter can, in general, have an arbitrarily form) :
\begin{equation} 
\psi'' + k^2 (z) \psi(z)=0, \ \ p(z)=\hbar k(z)=\pm\sqrt{2m[E-W(z)]}
\label{1}
\end{equation}
In the case of linear potential $W(z)=F z$ (that is, due to a constant force $F$), a mathematical equation is called the Airy equation. It can be written in a dimensionless form with a rescaling factor  depending only on $\hbar$, $m$ and $g$: $\tilde z_0=\hbar/(2m^3g)^{1/3}$. For the classical gravitational force in a vicinity of spherical source surface, the solution is interpreted as a QM 1D object called the quantum bouncer (QB). The QB is a pure academic concept used in toy model studies. There is a large literature relevant to the problem (\cite{QM} and elsewher); hence, our analysis is based on it. The Airy solution  $Ai(\xi)$ is described by quantum modes called the Airy functions with the corresponding eigenvalues (quantum states) $E_n=mgz_n$.  It is plotted in a dimensionless form in Fig. \ref{Airy}. 

\begin{figure}[t]
\includegraphics[scale=1.05]{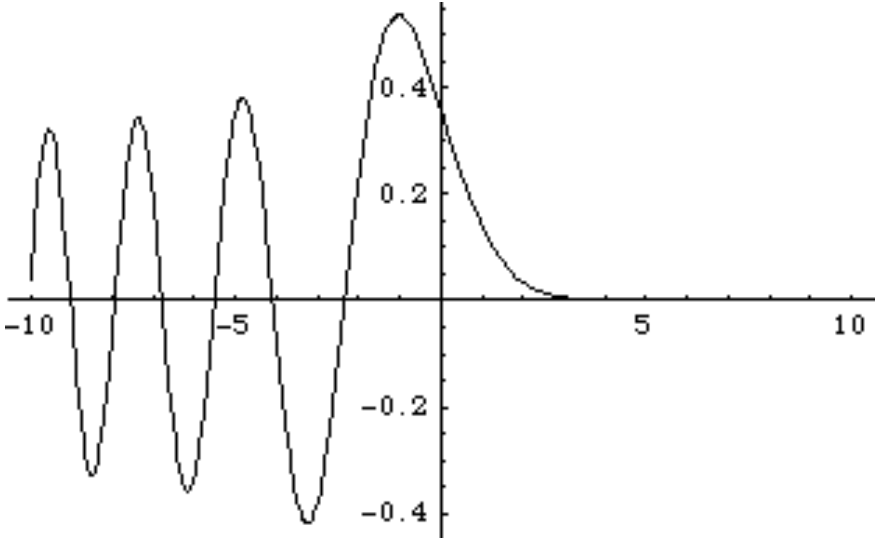}
\label{Airy}
\caption{\label{Airy}  The Airy function $Ai(s)$ for linear potential energy.  $s=\xi-\epsilon $, \ $\xi=(z/\tilde z_0)^{3/2}$,\  $\epsilon =E/mg\tilde z_0$,\  $\tilde z_0=(\hbar/2m^2 g)^{1/3}$. 
$Ai(s)\to 0$ as $s\to\pm\infty $}
\end{figure}

To create such a stationary object in laboratory is a great methodological challenge. The matter is that known existing QM bound systems are driven and stabilized by an internal (electromagnetic) field. They interact with external fields by exchanging resonant photons easily observed by spectroscopic methods, consequently, their nature is well understood within the quantum field theory. Contrarily, the QB is an example of a system driven by an external fields: the gravitational one and an elastic reaction of perfect mirror. Thus, the QB is an  academic non-radiating QM object without field theory extention.
 
Recently, results of measurements of ``gravitational quantum states'' of UCNs in the Earth gravitational field were reported in a series of publications [2-16]. The experiment was conducted at the reactor facility located at the Laue-Langevin Institute, Grenoble (further referred as the Grenoble experiment, by the authors, for short). 
The authors claim that quantum states in Earth gravitational field are observed for the first time. Though they do not directly use the term of ``QB states'', the states to be observed are numerically the QB exact ones. The matter is that a neutron above the mirror in the experiment is actually ``a running bouncer'' (that is, a 3D, at least 2D, in general, evolving quantum object). 

The present paper is devoted to a critical analysis of the experiment.


\begin{figure}
\includegraphics[scale=1.05]{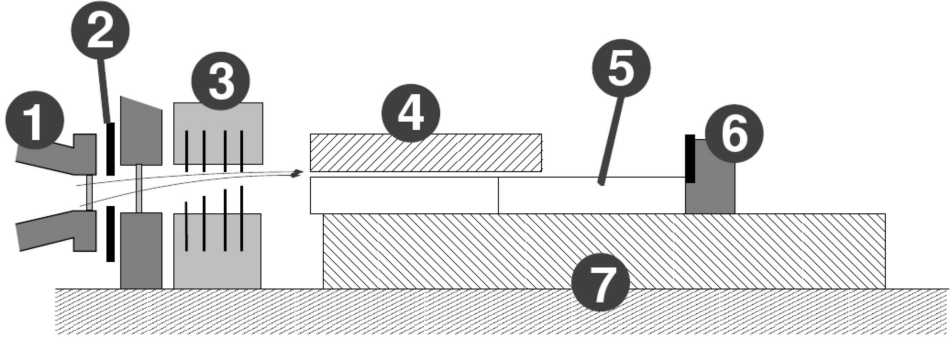}
\label{set}
\caption{\label{set} Experiment setup: interior of the vacuum chamber. Shown are the UCN guide (1);\   collimating blades (2);\   static collimator (3); \  the absorbing slit (4);\   bottom mirror (5);\   detector (6);\   mechanical support (7).\   Two possible trajectories of neutrons capable of entering the wave guide are shown  \cite{Nesvizh10}, \cite{Nesvizh18} }
\end{figure}


\section{Setup and general physical conditions}

The experiment setup  (see Fig. \ref{set}) has four parts:  the front part outside the slit, \-- neutron collimator and front aperture (1-3); the absorbing slit \-- absorption of ``unwanted neutrons'' and exit of  ``survivors'' (4); the bottom mirror including a part of the open mirror (outside the slit), where running and bouncing neutrons are supposed to be formed (5); the back part \-- detection of neutrons transmitted through the slit (6). 

In the experiment, a neutron in free fall between reflection points on the open mirror is studied on the subject of its gravitational quantum states. A neutron source is a high-power thermal nuclear reactor. There is a collimating UCN guide system to transport neutrons from the reactor to the horizontally placed, rectangular neutron wave guide (further called the slit). Incident neutrons flies into the slit through a front opening. They are characterized by a spectrum in the speed range of about $4-10$ $m/s$ and have a very small deviation from directionality of incidence. 

Since the incident neutron spectrum is not ``cold'' enough to provide UCNs in the ``quantization'' energy range, ``unwanted'' (high bouncing) neutrons should be removed. For this purpose, the upper wall of the slit is made of efficient absorber. Survived neutrons are those, which escape absorption and are able to pass through the slit. They come out of the slit through a back opening and become ''running bouncers'' that is, keep bouncing in a projectile motion in open space above the mirror before hitting an external detector (as shown in  Fig. \ref{set}). 

The absorber in the slit plays a role of a selector of neutrons having small vertical momenta $p_z<<p$, ($p$ is a magnitude of the total momentum). A  spectrum of $p_z$ depends on an absorber height $z_a$, while the latter is variable in the range of about $(1 - 60)\cdot 10^{-5}$ $m$.  The corresponding (expected) quantum levels in a vertical direction $E_n$ are in the range of few $peV$, roughly, in proportion $E_n=mgz_n$, where $m$ is neutron mass. Those energy values are about 5 orders less than actual kinetic energy $E_k$ of neutrons in the slit. A severity of experiment conditions is aggravated by non-coherence of the neutron source and the corresponding stochasticity of neutron events in the slit and the open mirror area. In particular, $E_k$ and ``a classical return point'' (or a hight of projectile trajectory) $z_{rp}$  are randomly distributed quantities in a wide range of their values. 

In an ideal classical picture, transmitted neutrons are those, which avoid absorption in a first reflection at a grazing angle $\theta\ge z_a /l_a$, where $l_a$ is a minimal distance between reflections (it is less or equal to the length of the absorbing slit). The angle can be assessed also as $\theta= p_z \ge p_{cr}$, were $p_{cr}$ is a critical value of $p_z$, corresponding to the critical height of the projectile trajectory $z_{cr} = z_a$. In the Grenoble experiment, the angle is of order $\theta \approx 10^{-4}$. 

Note.  ``The ideal picture'' is referred to the authors' {\em specularity}, or {\em  mirror perfectness},  assumption (equality of incidence and reflection angles in single events). A real mirror is characterized by ``grazing roughness''. Another cause of imperfectness may come from the fact that a mirror used in the experiment has a room temperature. 

In the QM approach, the amount of transmitted neutrons is less than the above (ideal) classical estimates because of Heisenberg's fluctuation of projectile heights in multiple reflections. Later we shall see that it makes a severe destructive effect:  at small $z_a$, the absorber ``kills'' neutrons having lowest $p_z$ value, a characteristic of the QB ground state in the authors' model.


The authors' plan of the experiment is based on the following premises and procedures: 

a)  Preparation, with the use of absorbing slit, of the assemble of neutrons having as low $p_z<<p$ as possible, and by this way provide conditions for quantum pattern formation in the $z$-distribution of probability density for survived neutrons (ones passed through the slit). In the open mirror area, they are supposed to become ``running bouncers'' characterized by the wave (Airy) function in the vertical direction. 

b)  Measurement of transmission curve (a count rate $T(z_a)$ as a function of absorber height $z_a$) with the following assessment of gravitational level parameters from measured data. The transmission function is thought to be associated with the $z$-distribution of neutron probability density $P(z, z_a)$ and the corresponding gravitational levels $E_n$ in terms of the above Airy function modes.

How those terms are related to the Airy equation in the authors' 1D model is seen from Fig. \ref{quantBouncer}, where the squared Airy functions $A_n (z, z_n)$ are plotted. They are proportional to the probability density $z$-distributions with respect to the mirror at $z=0$ position. Shown are states for $n=1,\ 2,\ 3,\ 4$.

It sould be noted that the resulting $z$-distribution of probability density $P(z, z_a)=\sum P_n(z, z_a)$ must look as a fairly smooth function: no distinct quantum pattern to be measured.



\begin{figure}[t]
\includegraphics{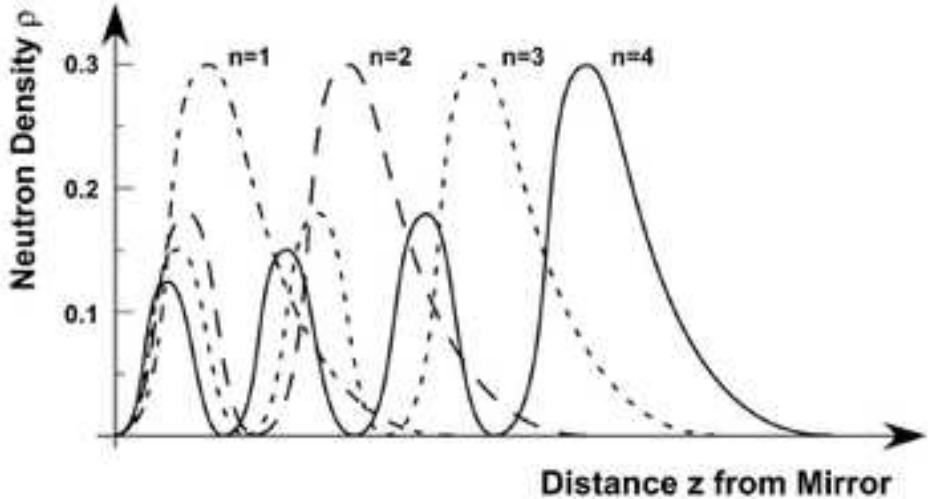}
\label{quantBouncer}
\caption{\label{quantBouncer} 
Squared wave functions, or probability density (not normalized) for four first states $E_n$: $N=1, 2, 3, 4$; bottom mirror position is fixed at  $z=0$; neutrons fall from distances in the upper half-space $0 \le z \le z_a$; $z_a$ \- \ the absorber height. By summing up the graphs, one is supposed to get a picture of resulting distribution $P(z, z_a)$ \cite{Nesvizh15}.}
\end{figure}

\section{Critique}

Thus, the purpose of the experiment is ``to observe'' running quantum bouncers by the detector placed at the end of the open mirror, while the detector counts {\em all} neutrons passed through the slit per second (as a function of the absorber height $z_a$). However, placing the detector at any distance from the back edge of the slit will not change the result (as far as emerged neutrons are not ``lost'' on their way to the detector). In particular, the detector of sufficient area can be placed somewhere behind the open mirror (outside the apparatus) or right at a back edge of the slit. In both cases, the Airy boundary conditions are not valid. It means that the measured transmission curve  $T(z_a)$ is not sensitive to a possible quantum pattern in the $z$-distribution of probability density  $P(z, z_a)$. Recall, the latter is defined in Airy function terms for a vertical motion of ``the running bouncer''. Obviously, counting neutrons at different $z_a$ is not ``the observation'' of their gravitational quantum states.

A count rate does not change either if the level of the open mirror is put lower than that inside the slit. We made this remark in connection with the authors' attempt ``to adjust'' the level, shifting the open mirror slightly lower (by about the first level distance $z_1=15$ $\mu m$) in order ''to pronounce the first step''  \cite{Nesvizh10}, \cite{Nesvizh18} (see the mirror cut in half by a black line in Fig.  \ref{set}). Not surprisingly, no effect was eventually found; so the ``adjustment'' idea was abandoned.

Therefore, we state that the authors' idea of extracting quantum information from measured transmission curve is the major methodological misconception: the treatment of relationship between $T(z_a)$ and $P(z, z_a)$ is wrong. Let us track the root of the misconception.

The authors seemingly start with a right approach when associate quantum levels with the probability density $z$-distribution for individual states $P_n(z, z_a)$ of a quantum bouncer running in the open mirror region. The ideal neutron assemble is characterized by a superposition of modes $n=1,\ 2... N(z_0)$, which determine the resulting probability density $P(z, z_a)$ (as in Fig. \ref{quantBouncer}). The problem begins when they introduce the concept of transmission function $T(z_a)$ to be measured and analyzed in connection with the Airy wave function and the corresponding  $P(z, z_a)$. 

The misconception is rooted in ``the principle of observation'' introduced by the authors  \cite{Nesvizh6},  \cite{Nesvizh7}, \cite{Nesvizh15}:

{\it ``Below about 15 $\mu m$, no neutrons can pass the slit... Ideally, we expect a stepwise dependence of transmission as a function of width. If the width is smaller than the spatial width of the lowest quantum state, then transmission will be zero. When the width is equal to the spatial width of the lowest quantum state, the transmission will increase sharply. A further increase in the width should not increase the transmission as long as the width is smaller than the spatial width of the second quantum state. Then again, the transmission should increase stepwise. At sufficiently high slit width one approaches the classical dependence.''}

Our comment on ``the principle of observation'' is, as next. As was explained, the transmission function $T(z_a)$ carries no information about physical processes occuring with neutrons in the open mirror area. The information, to which $T(z_a)$ is sensitive, solely concerns processes inside the absorbing slit. The transmission is proportional to a ratio $ N_a (z_a)/N_0  (z_a) \sim T(z_a) $, where $N_a (z_a)$ is a count rate of neutrons passed through an {\em absorbing} slit, and $N_0 (z_a)$ is a count rate of neutrons passed through a {\em non-absorbing} slit under all other equivalent conditions. The ratio is a probability  of neutron ``survival'' in a multiple reflections inside the slit,  $p_{sur} (z_a)$, which is a function of $z_a$. It depends on an average number of bouncings $n_b (z_a)$, or the length of absoring slit. 

Theoretically, the transmission $T(z_a)$ is a functional of a neutron wave function {\em inside the absorbing slit}. The functional is described by the non-stationary non-separable equation for the absorbing slit, the solution of which has nothing to do with the Airy function (since the absorber greatly influences the wave form and makes the mode essentially decaying). Roughly, $p_{sur} (z_a, n_b)\approx [1-p_a(z_a)]^{n_b}$, where $p_a(z_a)$ is a probability of absorption in one bouncing (strictly speaking, it is a function on $n-th$ bouncing). Once a neutron passed through the slit, its probability of being counted is fixed (ideally 100 \% ). 


To understand how the authors came up with the (criticized) ``stepwise'' picture, consider two, intuitively similar but physically different quantities: a neutron flux  $\phi$ measured and the probability density $P(z, z_a)$ looked for. One can mistakenly think that the quantum pattern in $P(z, z_a)$ is ``projected'' on the screen in a manner of wave diffraction picture from a coherent source. 

The actually measured quantity is a neutron count rate (the transmission 
$T(z_a)$)  
\begin{equation}
T(z_a)=\phi A (1-r) \epsilon  p_{sur}(z_a)
\label{2}
\end{equation}
where a flux $\phi=\rho v$ is a product of practically uniform quantities: a neutron density $\rho$\ $(particle/m^3)$ and a horizontal speed $v$; $A$ is a cross-section area of detected neutron flux, $\epsilon$ is a detector efficiency, and $r$ is a probability of reflection of incident neutron off the front aperture due to diffraction. In the experiment, $\epsilon$ and $(1-r)$ are close to $100$ \% , $A$ is proportional to $z_a$, and $v=\sqrt{v_x^2 + v_z^2}$  (the term $v_z$ must be disregarded as $v_z<<v$). Thus, the measured transmission is $T(z_a)\sim z_a p_{sur}(z_a)$. Neither $\rho$ nor $v$ are ``quantized''.

The authors' objective $P(z, z_a)$ is an analog of the classical probability density  $P_{cl}(z, z_a)\sim 1/v_z=dt/dz $, that has nothing to do with $\phi$ and $T(z_a)$. 
The method of ``gravitational level'' measurement must be analogous to the measurement of $P_{cl}(z, z_a)$ (the latter is not even mentioned in the authors' works).


The above discussed major misconception and other methodological deficiencies make the authors' claims not justified. Below we shall see that the experimental data themselves (''no steps observed'') are in contradiction with the authors' claim. We also note that some insignificant irregularities in measured curves could be instrumental ones that is, due to wave diffraction on the front aperture. Those (not discussed) instrumental effects could falsely imitate the physical ones, which the authors are looking for.

\section{Experimental results}



The experiment has a history of results published in a series of works, which we refer to ``early'' and ``latest'' periods. The two periods are principally different in key results of ``steps'' measurements, though the experimental setup has not been changed. Surprisingly, the ``steps'' were demonstrated in early works 2000-2004 \cite{Nesvizh21}, \cite{Nesvizh22}: 

{\em ``We report experimental evidence for gravitational quantum bound states of neutrons...  Under such  [slit] conditions, the falling neutrons do not move continuously along the vertical direction, but rather jump from one height to another, as predicted by quantum theory ''}. 
\medskip

The early measurements occurred to be technically or statistically mistaken. The authors tacitly admitted it in publications of new results and presented the ``improved'' version of experiment treatment and fitting procedure with some new phenomenological assumptions. 

A more accurately measured transmission function in Fig.\ref{UCNtransmLATER}, \cite{Nesvizh18}, appears a smooth curve, with no statistically significant ``quantum pattern''. The authors commented it:
 
{\em ``It was found, that except for the ground state, the stepwise increase is mostly washed out.''}

\begin{figure}
\includegraphics[scale=0.95]{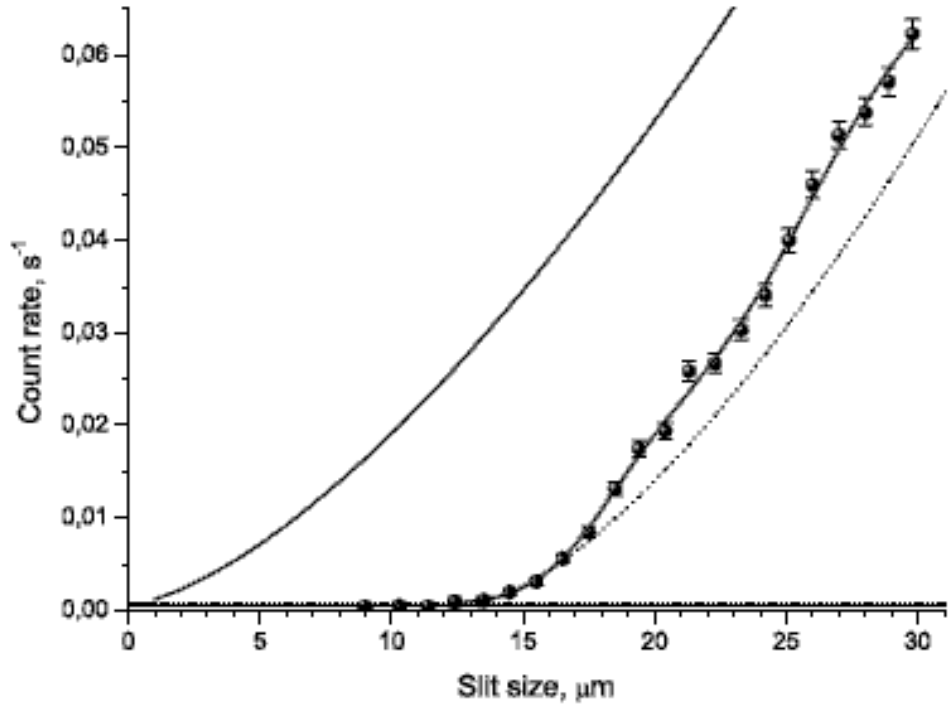}
\label{UCNtransmLATER}
\caption{\label{UCNtransmLATER} New measurements of neutron transmission: neutron counts versus rough scatterer/absorber height. 
The circles with error bars are the experimental data. The upper curve is a classical model $F_{cl}(z_a)\sim {z_a}^{3/2}$. The lower curve is a semi-classical model where only the lowest quantum level is taken into account. The experimental points are approximated by a model where higher levels are taken into account, \cite{Nesvizh18} }
\end{figure}








There is a ``cut-off'' in the curve: no counts at small absorber heights. The matter is that the absorber works as a neutron sink. At small heights, it ``kills'' neutrons in the ground state. Indeed, a bouncing (vertical) amplitude fluctuates in accordance with the Heisenberg uncertainty. For small widths  $z_a z_1 \approx z_a$, the uncertainty principle makes the effect: 
 $\Delta z_1 \Delta p_z \sim \Delta t_1 \Delta E_1  \sim h$, where $\Delta E_1 =mg z_1$.  In a course of multiple bouncing, a neutron eventually hits the absorber, what results in transmission blockage (a cut-off in the curve). To illustrate the problem, consider a neutron absorption probability $p_a (z_a)$ per one bounce at some absorber height $z_a$. Roughly, a quantum bouncer makes about 15 bounces in the slit. Hence, a chance to pass through the slit is $p_{sur}=(1-p_a)^{15}$. For some small small height, where $p_a=0.2$,\ the chance is about $p_{sur}= 0.035$. This is how a sharp cut-off is produced at about  $z_a= z_1$. After that, a transmission curve must be a monotonously increasing function. As noted, there could be some `irregularities'' in the curve due to residual effects of diffraction of neutrons on the front aperture.

In the final authors' claim, the statement of observation of  first (``ground'') state is emphasized. The authors admitted, however, that ``the population'' of the first level is inexplicably low. Indeed, how one can observe the ``ground state'' of neutrons, which are mostly ``killed'' by the absorber and for this reason do not exist in the open space above the bottom mirror, hence, are not detectable and, indeed, have not been detected.

\section{Conclusion}

The authors of the Grenoble experiment claim that neutron quantum levels (at least, the first one) in the gravitational field of Earth were observed for the first time. We consider such an experiment a great challenge because gravitational neutron quantum states were never observed before and the issue of their exitance is of great importance. Our detailed analysis of the experiment is focused on both methodology and measured data. Unfortunately, the conclusion is made that the objective and methodology of the experiment are not formulated in rigorous QM terms. The method of gravitational level measurement, in our view, is based on the misconception, hence, not adequate to the experiment objective. Our prediction that the measured data cannot reveal the quantum pattern is in an agreement with the actually measured data. 



\begin{thebibliography}{99}

\bibitem {QM} 

Leonard I. Schiff. ``Quantum Mechanics''. 2d Ed., McGraw-Hill Book Company (1955).

D. J. Griffins. ``Introduction to Quantum Mechanics'', Prentice-Hall, Enle-Wood Cliffs, New Jersey (1994). 

M. A. Doncheski, R. W. Robinett. ``Expectation value analysis of wave packet solutions
for the quantum bouncer: short-term classical and long-term revival behavior''. arXiv:quant-ph/0307046v1 (2003);

M. R. Brown. ``The quantum potential: the breakdown of classical symplectic symmetry and the energy of localization and dispersion''. arXiv:quant-ph/9703007v3  (2002);

Andreas Buchleitner, Dominique Delande, and Jakub Zakrzewski. ``Non-dispersive wave packets in periodically driven quantum systems''. arXiv:quant-ph/0210033v1 (2002).
 
J. J. Prentis, W. A. Fedak. ``Energy conservation in quantum mechanics''. Am. J. Phys., {\bf 72},  (2004) p. 580-590.
 
R. L. Gibbs. ``The quantum bouncer''. AJP, {\it 43}, 25-28 (1976).

J. Yoder. ``Using classical probability function to illuminate the relationship between classical and quantum physics''.    AJP, {\it 75}, 404-411 (2006).
  
Mario N. Berberan-Santos, E. Bodunov. Lionello Pogliani. ``Classical and quantum study of motion of particle in gravitational field''. J. Math. Chemistry, {\it 37}, 101-115, (2005).
 
R. W. Robinnett, M. A. Donchenski, L. C. Bassett. ``Simple examples of position-momentum correlated wave packet in 1D general form of time-dependent spread''. arXiv:quant-ph/0502097v1 (2005).

M. Belloni,. M. A. Donchenski, R. W. Robinnett. `` Exact results for ``bouncing Gaussianian wave packets''.
 arXiv:quant-ph/0408182v1 (2004).





\bibitem {Nesvizh1} H. Abele, S. Baessler, and A. Westphal. ``Quantum states of neutrons in gravitational field''.  arXiv: hep-ph/0703108v2 (2007). 


\bibitem {Nesvizh3} V.V.Nesvizhevsky, and K.V.Protasov. ``Quantum states of neutrons in the earth's gravitational field: state of the art, applications,perspectives''. Edited book on Trends in quantum gravity research. D. C. Moore. New York, Nova science publishers.  (2006) 65-107.

\bibitem {Nesvizh6} A. Yu. Voronin, H. Abele, et al. "Quantum motion of neutron in wave-guide in gravitational field". Phys. Rev. {\bf D73}, (4),  (2006) 044029. Also: arXiv: quant-ph/0512129 (2006).

\bibitem {Nesvizh7} A. Westphal, H. Abele, et al. "Quantum mechanical description of experiment on observation of gravitational bound states". arXiv: hep-ph/0602093 (2006). Also: Eur. Phys. J. {\bf C51}, 367-375 (2007).


\bibitem {Nesvizh9} A.E.Meyerovich and V.V.Nesvizhevsky. "Gravitational quantum states of neutrons in a rough waveguide."  arXiv: quant-ph/0603203 (2006).

\bibitem {Nesvizh10} C. Krantz. ``Quantum states of neutrons in the gravitational field''. Diploma thesis, (2006).


\bibitem {Nesvizh12} G. Pignol, K. V. Protasov, V. V. Nesvizhevsky. "A note on spontaneously emission of gravitons by a quantum bouncer". Class. Quantum Grav., (2007) {\bf 24}, 2439-2441.    


\bibitem {Nesvizh14} A.E.Meyerovich and V.V.Nesvizhevsky. "Gravitational quantum states of neutrons in a rough waveguide." Phys. Rev. {\bf A73},  (2006) 063616.


\bibitem {Nesvizh15}   H. Abele, S. Bassler, A. Westphal. ``Qantum states of neutrons in gravitational field and limits for non-Newtonian interaction in range between 1 and 10 micrometers''. arXiv: hep-ph/0301145v1 (2003).
Also: arXiv: hep-ph/0703108v2 (2007)


\bibitem {Nesvizh16} V.V.Nesvizhevsky, A.K.Petoukhov, et al. "Investigation of the neutron quantum states in the earth's gravitational field." Journal of Research of the National Institute of Standards and Technology {\bf 110} (3), (2005) 263-267.

\bibitem {Nesvizh18} V.V.Nesvizhevsky, A.K.Petukhov, et al. ``Study of the neutron quantum states in the gravity field''. Eur. Phys. J. {\bf C40}, 479 (2005). Also: arXiv: hep-ph/0502081 (2005). 

\bibitem {Nesvizh19} Nesvizhevsky VV, Protasov KV. "Constrains on non-Newtonian gravity from the experiment on neutron quantum states in the earth's gravitational field." Classical and Quantum Gravity {\bf 21},  (2004) 4557-4566.

\bibitem {Nesvizh20} V.V.Nesvizhevsky. "Quantum states of neutrons in the gravitational field and interaction of neutrons with nanoparticles." Uspekhi Fizicheskikh Nauk {\bf 46}, (1)  (2003) 93-97.

\bibitem {Nesvizh21} V.V.Nesvizhevsky, et al. "Measurement of quantum states of neutrons in the Earth's gravitational field." Phys. Rev. {\bf D67},  (2003) 102002. Also: arXiv: hep-ph/0306198 (2003) 

\bibitem {Nesvizh22} V.V.Nesvizhevsky, et al. "Quantum states of neutrons in the Earth's gravitational field." Nature, {\bf 415}, (2002) 297-299. 


 
\end{thebibliography}
\end{document}